# Multiple Spin State Analysis Applied to Graphite-like Carbon-based Ferromagnetism

Norio Ota[1], Narjes Gorjizadeh[2] and Yoshiyuki Kawazoe[2]
[1] Hitachi Maxell Ltd., Osaka, Japan   [2] Tohoku University, Sendai, Japan

Recent experiments indicate room-temperature ferromagnetism in graphite like materials [1-4]. This paper offers an multiple spin state analysis to find out the origine of ferromagnetism in case of nano meter size graphene molecule.

First principle density function theory calculation (DFT-GGA with 631-G basis set) is applied to nano meter size asymmetric graphene fifteen molecules. Major results are,

(1) Dihydrogenated zigzag edge molecule like $C_{64}H_{27}$ show that the most stable (lowest molecular energy) spin state is the highest one as $Sz=5/2$. Examples for spin density map of $Sz=1/2, 3/2$ and $5/2$ is shown in Fig.1. In other molecules like $C_{56}H_{24}$, $C_{64}H_{25}$, $C_{64}H_{22}$ and $C_{64}H_{23}$ also show the highest spin state most stable as shown in Fig.2. Energy difference between most stable spin state and next one overcome temperature difference 1000K, which suggests a stability of room temperature ferromagnetism.

(2) Radical carbon zigzag edge molecules are also analysed. As illustrated in Fig.3, in every five molecule, also the highest spin state is most stable.

(3) In contrast, nitrogen substituted molecules like $C_{59}N_5H_{22}$, $C_{61}N_3H_{22}$ etc. show opposite result, that is, the lowest spin state is most stable as shown in Fig.4.

There are following three key issues to bring those results.
(A) Edge specified localized spin arrangement [5-6].
(B) Up-Up (also Down-Down) complex spin pairs inside of molecule.
(C) Optimized atom position rearrangement depend on the spin state.
Detailed mechanism will be discussed in the Symposium. Multiple spin state analysis is very useful to design carbon based ferro-magnet and also to design new spintronic devices.

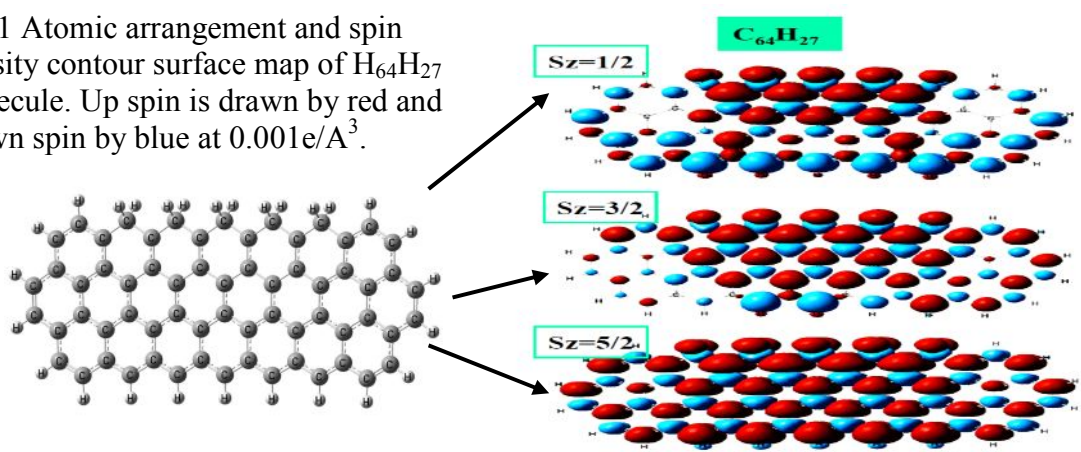

Fig.1 Atomic arrangement and spin density contour surface map of $H_{64}H_{27}$ molecule. Up spin is drawn by red and Down spin by blue at $0.001e/A^3$.

Fig.2 Total energy difference for dihydrogenated zigzag edge five molecules. Every molecule has several spin states with different total molecular energy.
In every molecule, the highest spin state is most lowest and stable energy state.

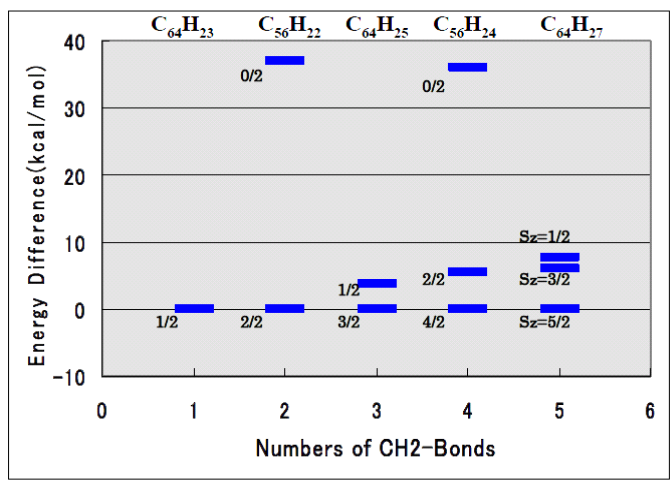

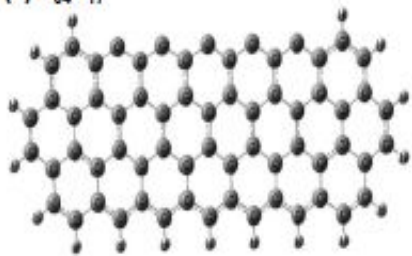

Fig.3 Energy difference between spin states for graphene molecule with radical carbon edges.
All molecules show that the highest spin state is most stable.

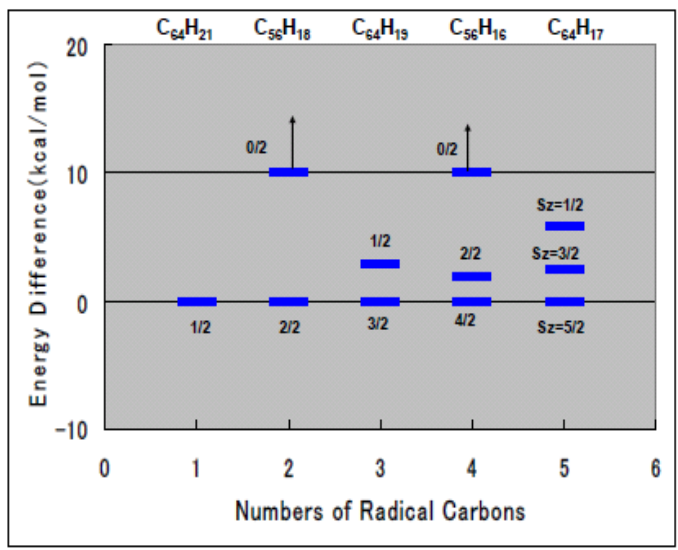

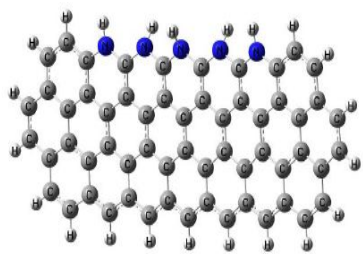

Fig.4 In case of Nitrogen substituted edge graphene molecule, most stable spin state is the lowest one like Sz=0/2 or 1/2.

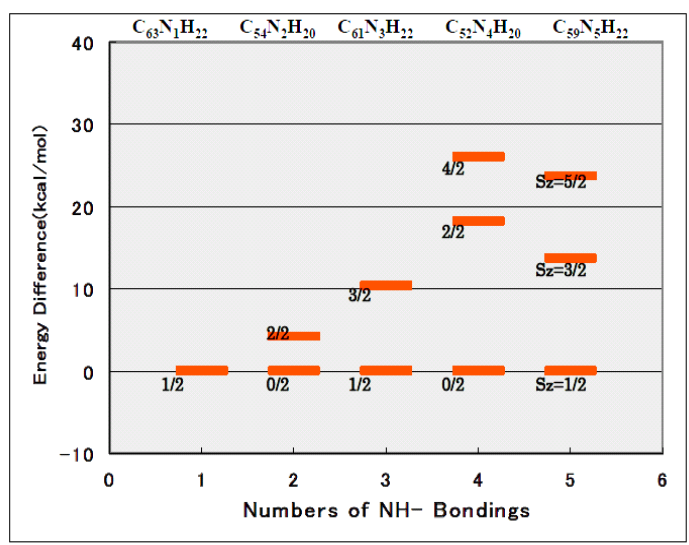